\documentclass[]{aa}
\usepackage{graphicx}
\begin{document}

\title{Possibility of a photometric detection of ``exomoons''
}

\titlerunning{Possibility of a photometric detection of ``exomoons''}

\author{
	Gy. M. Szab\'o,\inst{1,2},
	K. Szatm\'ary,\inst{1}
	Zs. Div\'eki\inst{1} \and
	A. Simon\inst{1}
	}

\authorrunning{Gy.M. Szab\'o et al.}

\offprints{Gy. M. Szab\'o}

\institute{
	Department of Experimental Physics \& Astronomical Observatory,
	University of Szeged,
	6720 Szeged, Hungary\\
	\email{szgy@titan.physx.u-szeged.hu}
   \and
	visitor at the Harvard-Smithsonian CfA, Cambridge, MA\\
	}

\date{}

\abstract
	{}
	{
We examined which exo-systems contain moons that
may be detected in transit.
	}
	{ 
We numerically modeled transit light curves of Earth-like and giant planets
that cointain moons with 0.005--0.4 Earth-mass. 
The orbital parameters were randomly
selected, but the entire system fulfilled Hill-stability. 
	}
	{
We conclude that the timing effect is caused by two scenarios:
the motion of the planet and the moon around the barycenter.
Which one dominates depends on the parameters of the system.
Already planned missions (Kepler, COROT) may be able to
detect the moon in transiting extrasolar Earth-Moon-like systems with a 20\% probability.
From our sample of 500 free-designed systems, 8 could be detected with
the photometric accuracy of 0.1 mmag and a 1 minute sampling, and one 
contains a stony planet. With ten times better accuracy, 51 detections
are expected. All such systems orbit far from the central star, with the orbital
periods at least 200 and 10 days for the planet and the moon, while
they contain K- and M-dwarf stars. Finally we estimate that a
few number of real detections can be expected by the end of the COROT and
the Kepler missions.
	}
	{}


\maketitle


\section{		Introduction}

The photometric transit detection of exoplanets has become an effective tool
in the past 6 years. The known systems (9 transits up to now) are 
summarized in the Schneider catalogue (2005).
In the future a few programs are planned to achieve measurements that are
as accurate as ca. 0.01--0.1 mmag, allowing the detection of a large sample of
Earth-like planets while they are in service (e.g. {\sc corot} mission, Auvergne et al.,   
2003,
Kepler mission, Borucki et al., 2003, Basri et al., 2005). The power of photometric detection
is expected to keep increasing in the future.

Sartoretti  \& Schneider (1999, SS99), Deeg (2002, D02), and Doyle \& Deeg 
(2003) argue that a larger moon than 
the Earth that is orbiting around an exoplanet causes measurable photometric
effects. Although there are only negative ground-based observations 
(Brown et al., 2001, Charbonneau et al., 2005), it is expected
that the planned space missions will be able to discover the exomoons.

\section{		Light curve simulations}

Light curve calculation consisted of modeling dynamically stable planet-moon (P--M) 
systems around real star models. The star models were based on the latest
Padova isochrones (Girardi et al., 2002), supplemented with
Phoenix linear limb darkening coefficients (Claret, 2000). Both the planet
and the moon had circular orbits, with the inclination taken into
account. The other input model parameters were (i) the age and suspected 
metallicity of the star, (ii) the orbital periods, masses, and radii of the
planet and the moon, and (iii) the photometric filter of the modeled observation.

So as to exclude the short-time escape of the moon, the system had to
fulfill Hill stability. According to it, the $\rm a_{moon}$ radius of the
moon was inside the L2 Lagrangian-point; and the $C_2$ Jacobi-constant at
L2 was smaller than for the orbiting moon, so the total
energy balance prohibited escaping through L2. Mathematically
\begin{equation}
C_{\rm moon} \equiv 2 \Omega_{\rm moon} - v^2_{\rm moon}> C_{\rm L2},
\end{equation}
where the effective potential 
$\Omega(x,y)$ denotes the same as is common for the circular restricted TBP,
$v$ is the velocity of the moon, and $C_{\rm L2}$ is simply $2 \Omega_{L2}$.

The first step in modeling was to determine the radius of the star from the
input parameters. Then the radii of orbits were calculated for the 
star--planet and the planet--moon systems according to Kepler's third law.
Although the exact solution of the moon's orbit requires solving a three-body
problem (TBP), we included an approximation when the planet and the
moon were orbiting with uniform velocity around their barycenter. This
approximation is good to the third order in the distance ratio (i.e. moon--planet
to planet--star) and in time, and is fully acceptable within the needs of
the main problem.

The apparent stellar flux during the transit was evaluated from a simulated
image of a real star model star in which pixels are zeroed by the transiting
objects. The stellar diameter was 1000 pixels, and fluxes were normalized to
out-of-transit stellar flux.

\begin{figure}
	\centering
	\includegraphics[width=8.2cm]{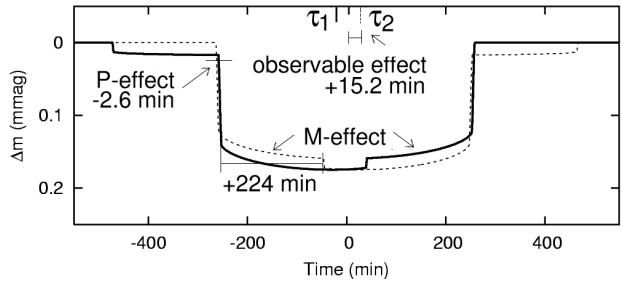}
	\caption{A sample light curve of two transits with a leading (solid
		line) and a trailing moon (dashed line) with the time differences. 
		The P-effect of $-2.6$ minutes is majorated by the 
		M-effect between the transits of the moon (224 min), while
		the observable effect is 15.2 min. 
		}
	\label{modelcurve}
\end{figure}

A sample light curve pair is presented in Fig. 
\ref{modelcurve}. The mass of the central star is 0.7 M$_{\sun}$, while the planet 
and the moon have similar sizes as the Earth and the Moon, respectively, the 
orbital velocity of the planet is 30 km/s, and the orbital period of the moon is
29 days.

When simulating { observations}, we sampled the previously
modeled light curves discretely, and co-added adjustable Gaussian noise. This
involved two more model parameters such as the sampling rate and 
photometric accuracy. We set them arbitrarily, and also included the expected 
data quality of the planned missions. This latter meant measurements whith 
0.1 mmag accuracy, sampled every minute. Although the standard sampling 
rate of Kepler will be 15 minutes, the transits will be read out 
once every minute for better resolution (D. Latham, private communication). 
The standard sampling of the {\sc corot} will be 8 minutes with the 
possibility of 32-second readout for each of 32 highlighted targets. 
We present two model observations in Fig. 2.

We calculated the timing effects according to the method of D02. 
Surprisingly, the method gave that the timing effect often appeared reversed 
(``earlier'' transit with leading moon) and with larger time shifts as 
predicted by the previous works cited above. The reason for this turned out 
to be a second scenario present in the timing effect, which we describe in the 
following.

\subsection{		Timing effects}

We define the $\tau$ central time of transit points as D02,
\begin{equation}
	\tau:=
	{\sum_{transit} ( t_i \cdot \Delta m_i)  \over \sum_{transit} \Delta m_i}\
\end{equation}
where $t_i$ and $\Delta m_i$ are the observational times and measured 
differential magnitude. The summation index $transit$ means that only 
the points belonging to the transit have to be taken into account. When the
photometric points are equally sampled, $\tau$ will be the time when the
planet passes before the central meridian of the star within the statistical
errors. (In case of non-uniform sampling, one has to use statistical weights, 
but in the case of the modeled automatic measurements the sampling is uniform.) 
Let $\Delta T=\tau_o - \tau_c$ mean the time difference between the observed 
and the expected central time (calculated if the planet had no moon)
of the transit.

In the presence of a moon the planet revolves around the 
barycenter, and some transits of the planet occur somewhat earlier or later 
than expected (SS99, D02). On the other hand,
the same is also true for the moon: depending on its trailing or leading position,
its individual transit occurs earlier or later. So as to distinguish those
scenarios, we will call the timings due to transit 
of the planet and the moon as P-effect and M-effect, respectively.

As the moon orbits farther from the barycenter, the M-effect can exceed
the P-effect in time by quite a lot, but as the moon is tiny, this effect is much less
in magnitude. In practice we observe some combinations of
P- and M-effects. 
If the moon is {leading}, the first half of the combined transit curve will be 
slightly deeper, and shallower after the moon finishes the transit (Fig. 1). 
The points referring to earlier times will get slightly more weight (Eq.
2.), and finally $\tau$ may refer to an {earlier} time. This is exactly
what we found in our model observations. Even when the moon was too tiny to cause
evident light curve distortions, $\tau$ could predict the presence of the moon,
due to the robust averaging in Eq. 2.

The effect that has more influence on the detected transit time depends on the
system parameters. If the moon is large and orbits far from the planet, 
the M-effect can far exceed the P-effect. For example, the P-effect is 
$-2.6$ minutes for the Sun-Earth-Moon 
system, its amplitude is 0.1 mmag (D02), the M-effect is 224 
minutes, and its amplitude is 0.0085 mmag. The observed effect will roughly 
the magnitude-weighted average, $(0.0085*224-0.1*2.6)/(0.1085)\approx 15.2$ 
minutes.

We note that the moon detection is more difficult 
if there is a large spot on the
stellar surface. The spot can cause light curve variations
thus $\tau$ may vary without the presence of a moon. 

\section{		Detectable systems}

\subsection{		The Earth--Moon-like systems}

Will the mission Kepler be able to find Earth-Moon-like systems around solar-like
stars? The photometric effect of the Moon is 0.0085 
mmag, so compared to the photometric accuracy of about 0.01--0.1
mmag the answer would be ``no''. But there is
hope for the detection of a Moon-sized moon around an Earth-sized exoplanet,
due to the combined effect exceeding 15 minutes. We simulated 0.1 mmag-quality     
measurements with 1-minute samplings. The observed scenario was similar to
Fig. 2, but with smaller amplitudes because the central star was 1 M$_{\sun}$. 
We calculated 20 events consisting of 4-4
neighboring transits, and randomized the initial orbital phase of the moon.
We found 5 detections in 20 calculations -- this detection
is not very probable, but cannot be excluded.

We defined a similar system in order to decide whether the photometric
accuracy or
the sampling rate is more dominant (Fig. 2) for a successful detection. 
We selected a  0.7 M$_{\sun}$ central star with 
$t=$5 Gyr and Z=0.019 metal content, while the planet and
moon masses, sizes, and periods
were the same as for the Earth-Moon. It was found that the success of 
the detection was
primarily determined by the sampling rate. Even with the 0.1-mmag photometric
accuracy, we got a 3$\sigma$ detection if the sampling rate was 1 or
2 minutes. In contrast, no positive detection was found with a 30-minute
sampling rate, and the positive detections with 10 and 20-minute sampling
rates were also rather ambiguous (Fig. 3).

\begin{figure}
	\centering
	\includegraphics[angle=270,width=8cm]{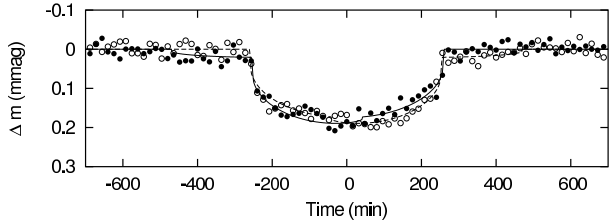}
	\caption{
		Simulated observations (0.1 mmag-quality, 1 min sampling,
		15 min averages for perspicuity) showing the 
		central transit of an Earth-sized planet and a Moon-sized 
		moon before a 0.7M$_{\sun}$ star (dots: leading moon, circles: trailing moon).
		}
	\label{earthcurve}
\end{figure}

\begin{figure}
	\centering
	\includegraphics[height=4.1cm]{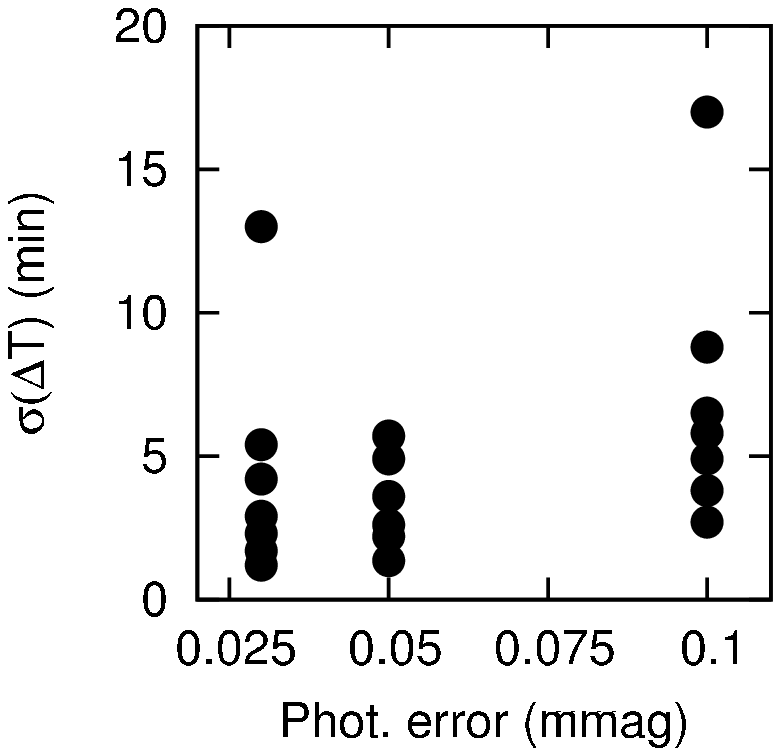}\ \
	\includegraphics[height=4.1cm]{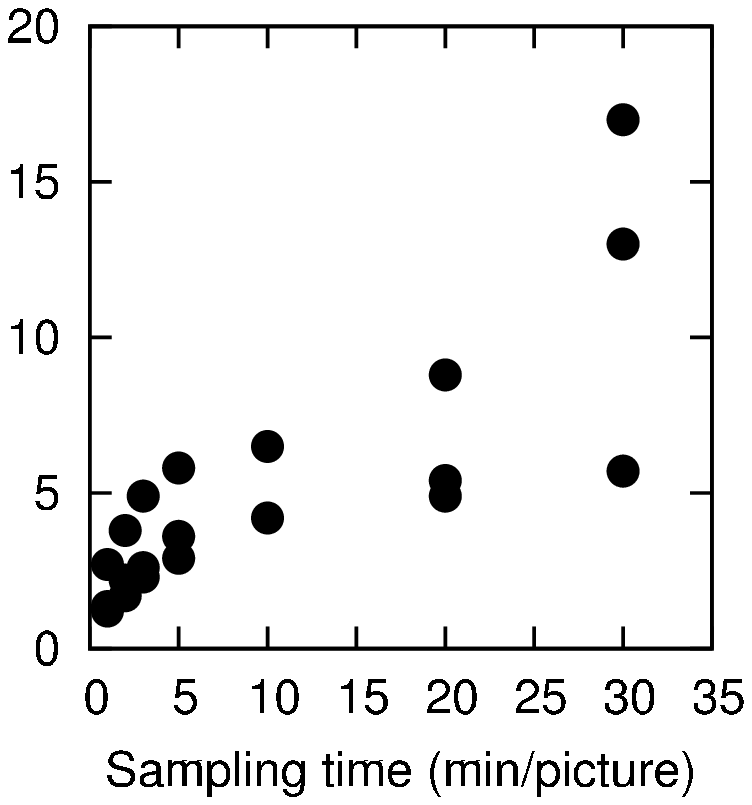}
	\vskip3mm
	\includegraphics[height=4.1cm]{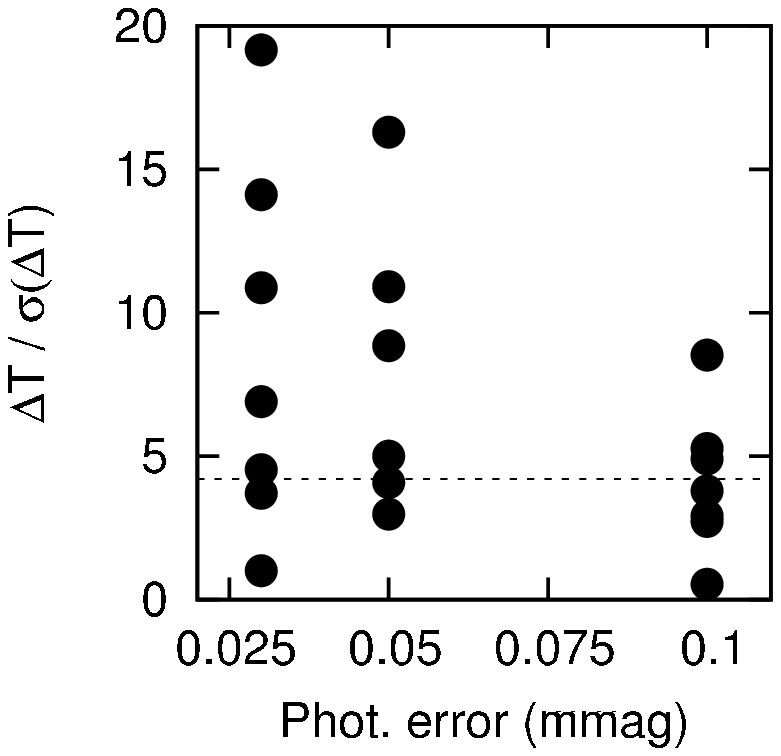}\ \
	\includegraphics[height=4.1cm]{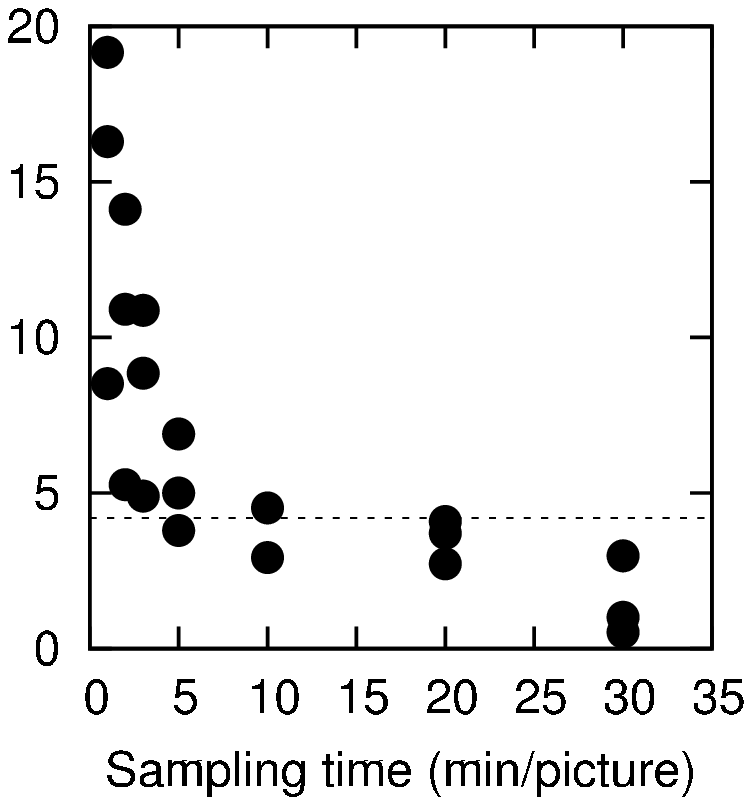}
	\caption{
		Uncertainties of the timing effect in the described 
		Earth--Moon-like system. Top: the panels show the 1$\sigma$ 
		ambiguities of the time delays as a function of photometric 
		accuracy (left panel) and of sampling interval (right panel). 
		Bottom: the same but expressed in relative errors; the
		simulations leading to 3$\sigma$ positive detections are above 
		the dashed line.
		}
	\label{eartherror}
\end{figure}

\subsection{		Other detectable systems}

\begin{figure*}
\centering
\includegraphics[height=5.15cm]{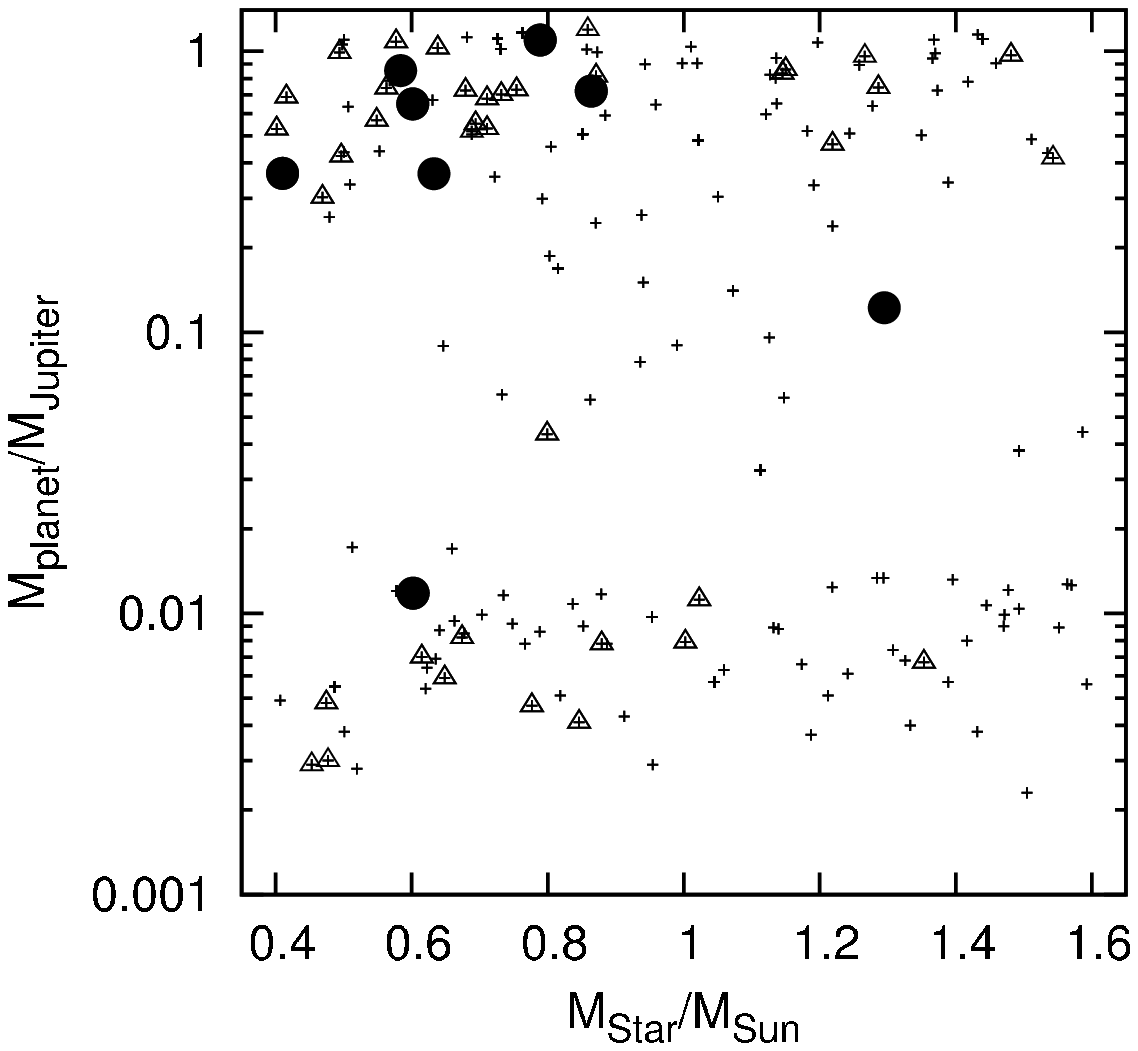}\ \
\includegraphics[height=5.3cm]{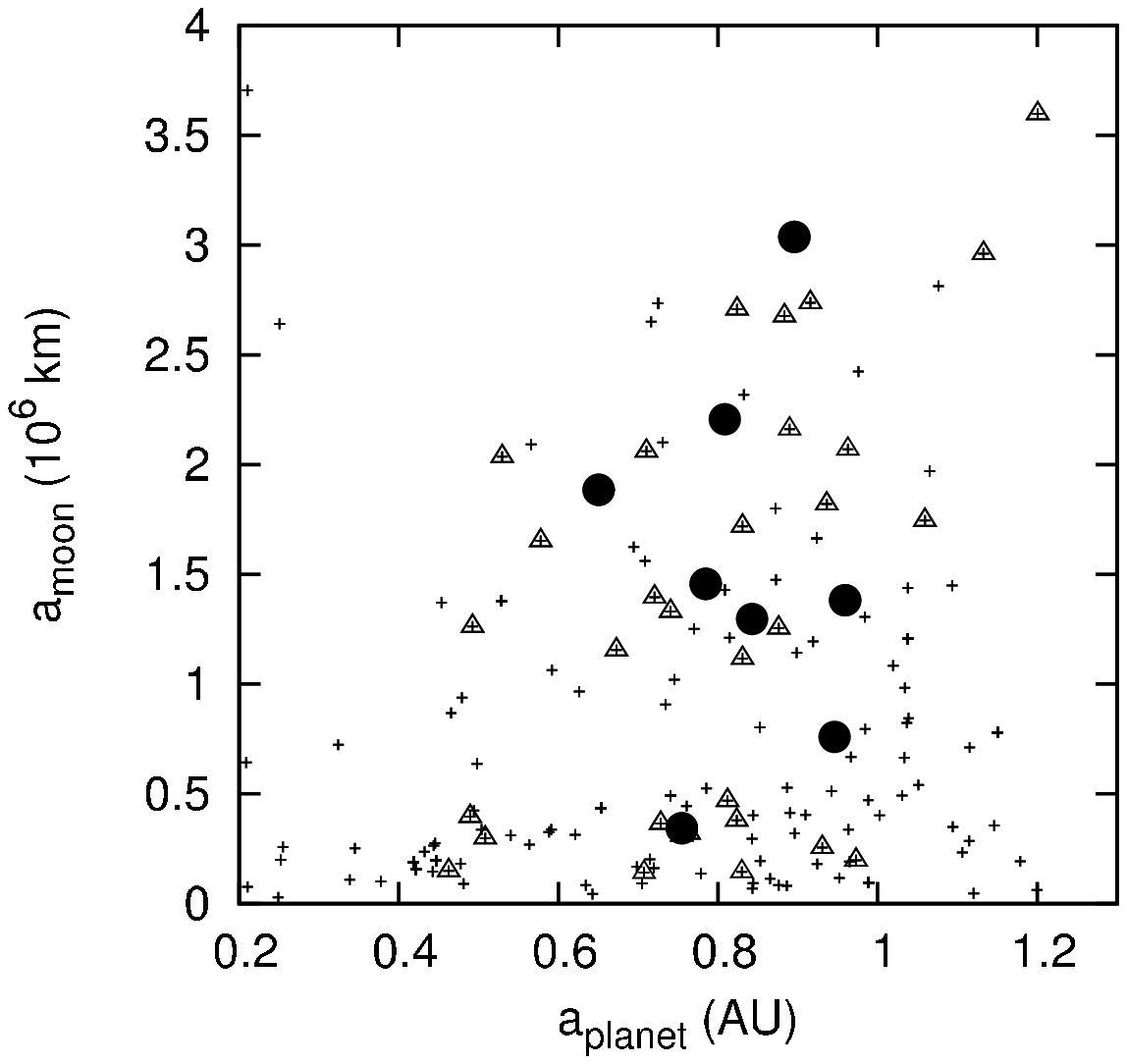}\ \ \ \
\includegraphics[height=5.2cm]{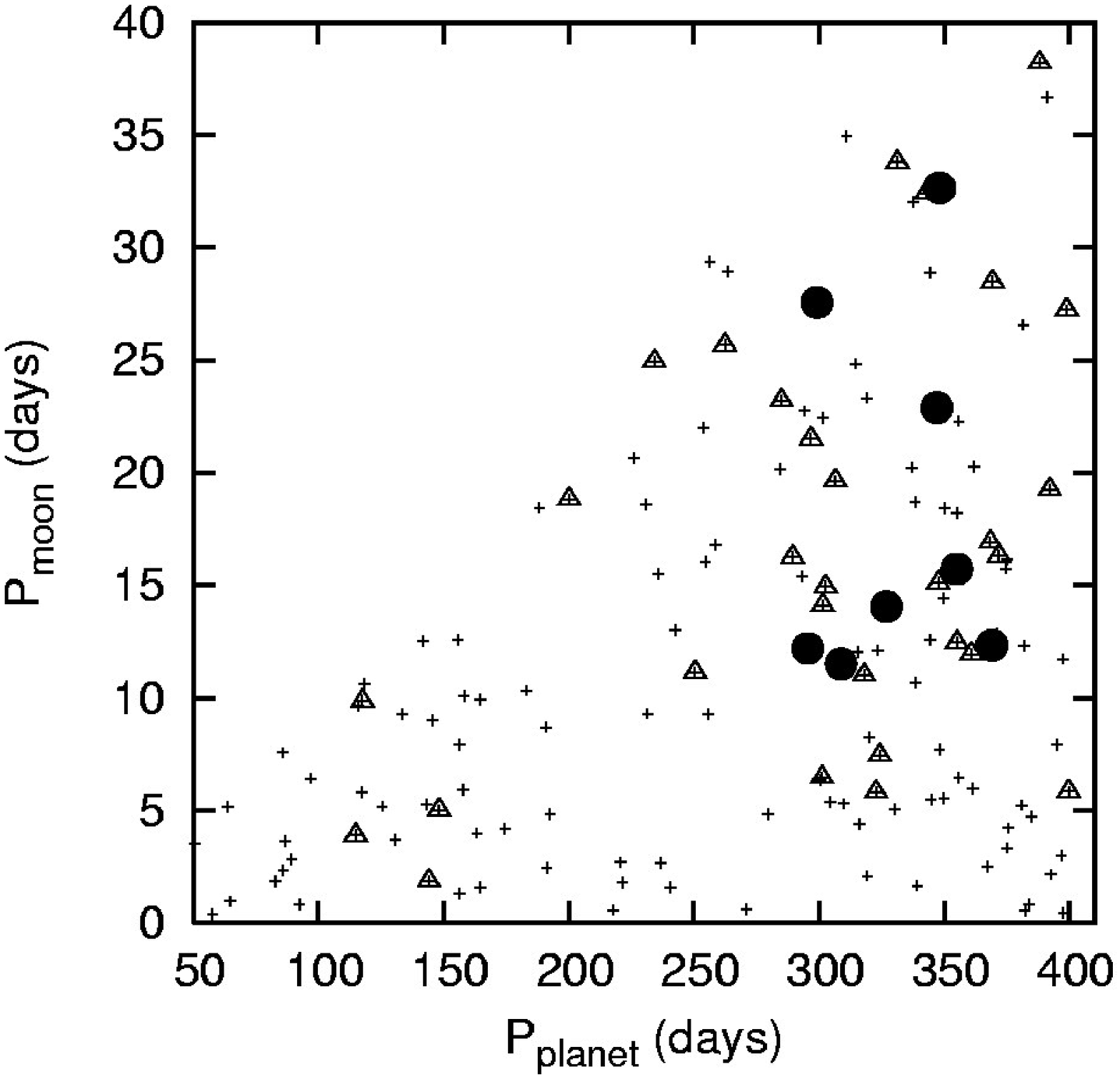}
\caption{
Systems with a detectable moon, if the photometric accuracy is 0.1 mmag (large
dots) or 0.01 mmag (triangles); undetectable systems are marked by small crosses.
}
\label{allsim}
\end{figure*}

From an observational point of view, it is interesting to
characterize systems that contain observable moons. We applied a
Monte-Carlo method for finding some appropriate systems. We randomly
generated systems, simulated the transits, and checked if the
moon is observable with at least 3$\sigma$ confidence. Five hundred systems 
were modeled containing both giant and Earth-like planets, different
inclinations, planet and moon densities, and orbital periods, where 
the period of the planet did not exceed 400 days in
order to have at least 4 transits during the planned operation time of Kepler.
One set of model measurements were calculated with the ``present'' (0.1
mmag) accuracy with a
1-minute sampling rate, and a ``future observable'' set was included with 
ten times better accuracy (0.01 mmag).

We found 51 ``future observable'' systems and 8 that could be
observed with the ``present'' accuracy. Both giant and Earth-like planets can have
observable moons; in the first case the S/N of the transit is high enough to
allow accurate measurement. In the second case the combined effect 
exceeds a few (5-55) minutes as the masses of the moons were comparable to the
mass of the Earth-like planet. The present equipments allow the detection of
5 moons of giant planets around red dwarf stars, but a positive detection
suggests that there is a chance in the case of Earth-like planets. The planet
has to be at least 0.6 AU from the star when observing with the present 
accuracy, and at least 0.4 AU in the case of the ten-times accurate measurements. 

The present
equipments allow detection of the moons of giant planets around red dwarf 
stars (Fig. \ref{allsim}.,
left panel), but positive detections suggest that there is some chance in case of Earth-like planets, 
or even detections for G-stars. The ``future observable'' systems contain more detections with
an Earth-like planet,
while they seem to allow more positive detections for G- and K-stars, too.
There is also an upper limit to the photometric accuracy: when the modelled 
light curves had 0.15 mmag accuracy and so a bit less S/N, only 3 moons were 
positively detected. 

The middle panel confirms another important restriction: the semi-major axis of the planet has to be large
enough (0.6 AU with 0.1 mmag-quality, 0.4 AU with 0.01 mmag). 
The planets farther from the central star move more slowly, and they can have more 
distant Hill-stable moons. Both make detection easier.
Experimentally, almost
every third test system was positively detected where the moon was farther
than one million km from the planet (referring to the 0.01-mmag simulated 
accuracy). 

The right panel expresses the same in terms
of orbital periods: the planet must have at least about 280 days orbital period for promising detections. This
is not too favorable as one can expect only a few transits during a 4-year mission,
thereby reducing the chance for a somewhat detailed modeling of the moon itself.

This condition makes the detection of a moon around the well-known ``hot-Jupiter''
planets quite unlikely. Those planets revolve so fast that the combined effect
lies within the magnitude of
a second. The only way to make a detection would be with the shape variation:
for the smallest (0.7 R$_{\sun}$) central star, {\sc ogle-tr}113, the photometric effect of a Ganymede-sized
moon is less than 0.03 mmag, and only 0.2 mmag for an Earth-sized moon.
Their Hill-radius is rather small, extending only to twice the radius of
the planet, so the probability of a moon being present is low. Considering
their extended atmosphere as well, which may decelerate 
an orbiting body, the ``lifetime'' of any moon seems to be limited.

\section{Conclusions}

What is the suggested observational strategy in order to find exomoons? Although
the required photometric accuracy (0.1--0.15 mmag) seems to be about the
highest quality we can expect
nowadays, short sampling intervals should be
used whenever possible. This will help by increasing the number of systems where the timing
effect is less, e.g. when the moon is smaller, closer to the planet, or
the mass ratio is smaller. The strategy of the planned missions is concordant with this.

We note that the observing strategy of Kepler (Basri et al., 2005) may allow the
detection of exomoons, but may also lead to the rejection of those observations. 
The transit pipeline of Kepler includes at least 3 transits 
with timings that are consistent with a
periodically revolving planet, within the observational errors. Any
inconsistent transits will be rejected.
If an exomoon has an observable timing effect such that timings occur
{\it discordantly} under the assumption of strict monoperiodicity, 
Kepler may not recognize the light variation as a transit. 
Therefore we suggest including some timing variation in the acceptance
criteria.

The possible spectroscopic confirmation of a moon could be either the
observation of the perturbations in the radial velocity of the star or
the detection of the moon within the Rossiter-McLaughlin effect, which e.g.
Bouchy et al. (2005) observed for a transiting planet. Unfortunately 
both mean only slight variations in the fine structures, 
which is why there is really little hope for their success.

Based on the presented calculations, one may estimate the magnitude of the expected real
detections with Kepler. The total number of Earth-sized planets to be
discovered is a few hundred during the entire mission. If only 5 percent
of them have a moon similar to our Moon, and only every fifth moon are
really found, we should get a few positively detected exomoons. If we
take the {\sc corot} mission and the forthcoming missions into account, we
may have a few dozen known exomoons by the end of the following decade.

\section{Acknowledgements}
The research was supported by Hungarian OTKA Grant T042509.
We thank D. W. Latham for his hospitality and for the fruitful
discussions.

\end{document}